\tolerance=10000
\input phyzzx


\REF\roz{M. Rozali, Phys. Lett. {\bf  B400} (1997) 260,
hep-th/9702136.}
\REF\brs{M. Berkooz, M. Rozali, and N. Seiberg,
Phys. Lett. {\bf B408} (1997) 105,
hep-th/9704089.}
\REF\note{N.Seiberg, hep-th/9705117.}
\REF\hen{X. Bekaert, M. Henneaux and A. Sevrin, Phys. Lett. {\bf
B468} (1999) 228; hep-th/9909094.}
\REF\strom{A. Strominger, Phys. Lett. {\bf B383} (1996) 44, hep-th/9512059;
P.K. Townsend, Phys. Lett. {\bf B373} (1996) 68, hep-th/9512062.}
\REF\WitUSC{E. Witten, in Proceedings of Strings 95,
hep-th/9507121.}
\REF\sei{N. Seiberg,
Phys.Lett. {\bf B408} (1997) 98, hep-th/9705221.}
\REF\dvv{R. Dijkgraaf, E. Verlinde, and H. Verlinde,
hep-th/9603126;
 hep-th/9604055.}
 \REF\lit{O. Aharony, hep-th/9911147.}
 \REF\Cremmer{E. Cremmer, in {\it Supergravity and Superspace},
S.W.
Hawking and
M. Ro\v cek, C.U.P.
Cambridge,  1981.}
\REF\HT{C.M. Hull and P.K. Townsend, Nucl. Phys. {\bf B438} (1995)
109;  hep-th/9410167.}
\REF\Witten{E. Witten, Nucl. Phys. {\bf B443} (1995)  85,
hep-th/9503124.}
\REF\HStr{C.M. Hull, Nucl. Phys. {\bf B468} (1996) 113,
hep-th/9512181.}
\REF\matten{O. Aharony, M. Berkooz and N. Seiberg,
Adv. Theor. Math. Phys. {\bf 2} (1998) 119;
hep-th/97012117.}
\REF\seig{N. Seiberg, Phys.Lett. {\bf B390} (1997) 169; hep-th/9609161.}
\REF\GravDu{C.M. Hull, Nucl. Phys. {\bf B509} (1997) 252,
hep-th/9705162.}
\REF\falg{C.M. Hull, hep-th/0004086.}
\REF\julia {B. Julia in {\it Supergravity and Superspace}, S.W.
Hawking and M.
Ro\v cek, C.U.P.
Cambridge,  1981. }
\REF\GHaw{G.W. Gibbons and S.W. Hawking, Phys. Lett. {\bf 78B}
(1978)
430.}\REF\GP{G.W. Gibbons and M.J. Perry, Nucl. Phys. {\bf B248}
(1984)
629.}
\REF\SGP {R. Sorkin, Phys. Rev. Lett. {\bf 51} (1983) 87 ; D.
Gross
and
M. Perry, Nucl. Phys. {\bf B226} (1983) 29.}
\REF\HTE{C.M. Hull and P.K. Townsend, Nucl. Phys. {\bf B451}
(1995)
525,
hep-th/9505073.}
\REF\Strath{J. Strathdee, Int. Jour. Mod. Phys. {\bf A2} (1987)
273.}
\REF\ton{P.K. Townsend, Phys. Lett. {\bf 139B} (1984) 283.}
\REF\anom{E. Witten, J. Geom. Phys. {\bf 22 } (1997) 103, hep-th/9610234;
L. Bonora,  C.S. Chu,  M. Rinaldi, JHEP {\bf 9712 } (1997) 007, hep-th/9710063;
D. Freed,  J. A. Harvey,  R. Minasian,  G. Moore, Adv. Theor. Math. Phys. 2
(1998) 601, hep-th/9803205.
}


%

%
\font\mybb=msbm10 at 12pt
\def\bbcc#1{\hbox{\mybb#1}}
\def\Z {\bbcc{Z}}
\def\R {\bbcc{R}}

\def \aa {\alpha}
\def \bb {\beta}
\def \gg {\gamma}
\def \dd {\delta}
\def \ee {\epsilon}

\def \kk {\kappa}
\def \ll {\lambda}
\def \mm {\mu}
\def \nn {\nu}

\def \rr {\rho}
\def \ss {\sigma}
\def \tt {\tau}

 \def \ggg {\Gamma}

\def \www{\Omega}

\def \ti {\tilde}

\def \2 {{1 \over 2}}
\def \3 {{1 \over 3}}
\def \4 {{1 \over 4}}
\def \5 {{1 \over 5}}
\def \6 {{1 \over 6}}
\def \7 {{1 \over 7}}
\def \8 {{1 \over 8}}
\def \9 {{1 \over 9}}
\def \00 { \infty}

\def\++ {{(+)}}
\def \- {{(-)}}
\def\+-{{(\pm)}}

\def\ek {\eqn\abc$$}

\def \pa {\partial}


 \def\unit{\hbox to 3.3pt{\hskip1.3pt \vrule height 7pt width .4pt
\hskip.7pt
\vrule height 7.85pt width .4pt \kern-2.4pt
\hrulefill \kern-3pt
\raise 4pt\hbox{\char'40}}}

\def\nup#1({Nucl.\ Phys.\  {\bf B#1}\ (}

\def \qq {\qquad}


\Pubnum{ \vbox{ \hbox {QMW-00-03} \hbox{hep-th/0004195}} }
\pubtype{}
\date{April, 2000}

\titlepage

\title {\bf  Strongly Coupled Gravity and Duality }

\author{C.M. Hull}
\address{Physics Department,
Queen Mary and Westfield College,
\break
Mile End Road, London E1 4NS, U.K.}
\vskip 0.5cm

\abstract {
A strong coupling limit of  theories whose low-energy effective
field theory is 5-dimensional $N=8$ supergravity is proposed in
which the gravitational coupling becomes large.
It is argued that, if this limit exists, it should be a
6-dimensional theory with (4,0) supersymmetry compactified on a
circle whose radius gives the 5-dimensional Planck length.
The sector corresponding to the  $D=5$ supergravity multiplet
is a (4,0) $D=6$ superconformal field theory
based on the (4,0)
multiplet
with 27 self-dual 2-forms, 42 scalars and, instead of a graviton,
a fourth-rank tensor gauge field satisfying a self-duality
constraint. The superconformal field theory has 32 supersymmetries and 32
conformal supersymmetries and its
dimensional reduction    gives the
maximal supergravity in five dimensions. Electromagnetic duality generalises to
a gravitational triality. }

\endpage

\chapter{Introduction}

The concept of a space-time manifold as the arena for
physics may
have to be modified at some very small distance
scale, $L$, but at larger scales it should give
a good description. There are a number of scenarios
in which
gravity could become strong at some much larger length scale, $l
>>L$.
For example, in M-theory $L$  might be the 11-dimensional
Planck length
while $l$ could be  the length scale $l=\kk ^{2/D-2}$ giving the
gravitational coupling constant
$\kk$ for some $D$-dimensional gravity theory arising from a
compactification or brane-world
scenario.
At energy scales $E$
such that $1/l << E<<1/L$, gravity
is strongly coupled but one might still have
some sort of  conventional space-time description, and in
particular
there
could be a dual field theory
describing this strong coupling phase.
Alternatively, it could be that there  is some
dual theory describing the gravitational theory at
{\it all} energies above some lower bound $E>1/l$
without any cut-off at a higher scale.
In M-theory, there would be  dual \lq phases' of M-theory,
one of which provided a good description at low energies and the
other in a certain high energy regime, and each would have its
own field theory limit.

The aim is then to find a gravitational analogue
of some of the weak-strong coupling dualities that
have been found in field theories and string theories.
At first sight, it would seem that such  dual theories of gravity
would be
unlikely to exist, but the close relations between gravity and
gauge
theories and the implications such a dual
description could have for  quantum
gravity suggest that it could be worthwhile to investigate this
possibility.

A useful example to try and generalise is
 5-dimensional maximally supersymmetric Yang-Mills
theory. It is non-renormalisable and so
new physics should emerge at short distances.
Its strong coupling limit is a 6-dimensional (2,0) supersymmetric
theory,
compactified on a circle,
in which the gauge field is replaced by a 2-form gauge field with
self-dual field strength,
and the 5 scalar fields are all promoted to scalar
fields in 6 dimensions [\roz-\note].
The $D=6$ theory is believed to be
a non-trivial superconformally invariant quantum theory [\note]
and the $D=5$ gauge coupling $g_{YM}$
arises as the radius of the circle, $g_{YM}^2= R$.
The relationship between the $D=5 $ and $D=6$ theories is
straightforward to establish for the free case  in which
the Yang-Mills gauge group is abelian, but in the interacting
theory
the 6-dimensional origin of the $D=5$ non-abelian interactions
is mysterious; there are certainly no
local covariant interactions
that can be written down that give Yang-Mills interactions when
dimensionally
reduced [\hen].
Nonetheless, the fact that these $D=5$
and $D=6$ theories arise
as the world volume theories of D4 and M5 branes respectively   [\strom]
gives
strong support for the existence of such a 6-dimensional origin
for the gauge
interactions. The (2,0) conformal theory also arises in the compactification of
the IIB string on $K3$ [\WitUSC] and as the
zero-slope limit of a little string theory [\sei-\lit].

Interactions of the \lq mysterious' type that are
believed to occur in the (2,0)
conformal field theory will be referred to here as \lq
M-interactions'.
Given that such  interactions
with no conventional field theory formulation
arise in the M5-brane
world-volume theory, it is natural to ask whether
similar M-interactions could arise elsewhere;
given that they arise in one corner of M-theory, it seems
reasonable
to expect that there are other corners in which such phenomena
occur.

The theory that will be considered here
is five-dimensional $N=8$ supergravity (ungauged),
which has a global $E_6$ symmetry and a local $Sp(4)=USp(8)$
symmetry [\Cremmer].
It is non-renormalisable, and will be regarded as arising as a
massless sector
of some consistent theory, such as
M-theory compactified on a 6-torus, in which the global $E_6$
symmetry is broken to a discrete U-duality subgroup [\HT].
The massless bosonic fields consist of a graviton,
27 abelian vector fields and 42 scalars.
The action
is
$$S=\int d^{5}x \sqrt{-g}\left( {1\over l^{3}}R-
{1\over 4l^{}}F^{2}+\ldots \right)
\eqn\act$$
where $l=\kk ^{2/3}$ is the 5-dimensional Planck length.
If this does have a  dual at strong gravitational coupling, i.e. a
limit as
$l\to \infty$, then
it is natural to look
first for a theory with 32 supersymmetries and
with $E_{6 }\times Sp(4)$ symmetry, as
the  simplest possibility would be if these symmetries
survived at strong coupling.
The arguments used in this paper will be similar to those used in
  [\HT,\Witten,\HStr] for extrapolating to large values of
dimensionless
  string couplings and moduli, and
it will be assumed that all symmetries are preserved
and BPS states are protected and survive as the coupling $l$ is
increased.
The symmetries then impose stringent constraints on what can
happen,
and
one can then seek checks on the predicted strong-coupling  dual.

Comparison with the $D=5$ gauge theory case suggests seeking
a 6-dimensional theory
compactified on a circle in which the 27 $D=5$ vector fields
are replaced
with
27 self-dual 2-form gauge fields in $D=6$.
Indeed, the decomposition of $ N=8$ supergravity into $N=4$
multiplets
includes five $N=4$ vector multiplets with coupling
constant $g_{YM}^{2}=l$, and the strong coupling limit of each should give a
(2,0)
6-dimensional
tensor multiplet, as will be argued  in section
2.
However, the 27 vector fields of the $N=8$ theory fit into an
irreducible representation of $E_{6}$, and so if $E_{6}$ symmetry
is
to be maintained and if    some of the vector fields become
self-dual 2-form gauge fields in 6 dimensions, then all of them
should.
Furthermore, all the fields fit into an irreducible multiplet of
the
$N=8$ supersymmetry
algebra, so that
if the 32 supersymmetries survive at strong coupling and some of
the
fields become 6-dimensional, then the whole theory should become
6-dimensional.
As in the gauge theory case, all the scalar fields should survive
at
strong coupling and so should lift to
   42 scalars in 6 dimensions.
If the $D=6$ theory were conformally invariant, then
the $D=5$ gravitational coupling $l$ could
arise from the radius of the circle $R$.

This then suggests that the strong coupling dual field theory
should
be
 a superconformal theory in six dimensions
with 32 ordinary supersymmetries (and   a further 32 conformal
supersymmetries),  and should have
42 scalars and 27 self-dual 2-form gauge fields. The unique
supergravity theory in 6 dimensions with 32 supersymmetries
fails on every one of these requirements with the obvious
exception
  of having
32 supersymmetries.
Remarkably, such a conformal supermultiplet in six dimensions
with 32 supersymmetries, $Sp(4)$ R-symmetry, 42 scalars and 27
self-dual 2-form gauge
fields {\it
does } exist and
so is an immediate candidate for a strong coupling
dual. However, instead of a graviton, it has an exotic
fourth-rank tensor gauge field satisfying a self-duality
constraint.
The relation between
linearised 5-dimensional $N=8$
supergravity and this free 6-dimensional theory
will be established here,
but there are no covariant local interactions in $D=6$
that could give rise to
the $D=5$ supergravity interactions, so
that
their 6-dimensional origin
would have to be
some form of M-interactions. In this case
there is no analogue of the M5-brane
argument to support this,
although the M5-brane case does set a suggestive
precedent.
The purpose of this  paper is to
investigate the possibility that
the strong coupling limit of the $D=5$ theory is this
exotic 6-dimensional theory.
In the context in which the $D=5$ supergravity is the massless
sector
of M-theory on $T^{6}$, then the $D=6$ superconformal theory would be
a field theory sector of
a new 6-dimensional phase of M-theory.

\chapter{$D=5$ Super-Yang-Mills Theory and the $D=6$ (2,0) Tensor
Theory}

The $N=4$ Yang-Mills theory in five dimensions
has a vector field $A$, 5 scalars $\phi^I$ and
four 4-component  fermions $\ll^a $ satisfying a symplectic
Majorana
reality condition, all taking
values in the adjoint of the gauge group. There is a global
$Spin(5)=Sp(2)$
R-symmetry under which
the scalars transform as a {\bf 5} and the fermions as a spinor
{\bf
4}.
The action is
$$S=
-{1\over g_{YM}^2}
\int d^5
x \left(  \4 F^2 + \2 (D\phi)^2 +...
\right)
\eqn\ymact$$
The coupling constant $g_{YM}$
is dimensionful and $l=g_{YM}^2$ is a  length scale.
The theory is non-renormalisable, and so new physics is expected
at
energies
$E\ge 1/l$. The 5-dimensional gauge theory is then a low-energy
effective theory and it will be supposed that the $D=5$
super-Yang-Mills multiplet arises as
part of a consistent theory that can be extrapolated to   scales
$E>1/l$.
For example, it could arise as part of
the 5-dimensional heterotic string, or  on a stack of D4-branes
in the IIA string, and in both these cases there would be in
addition
supergravity fields and an infinite number of massive fields in
the
full theory.

Consider the behaviour of the theory as the coupling $g_{YM}$ is
increased.
The simplest possibility would be for the theory to continue to be
invariant under
16 supersymmetries and  the  $Spin(5)$ R-symmetry, for all 5 scalar
fields to remain  so that the theory has the same moduli space,
and
for the BPS
spectrum to remain the same, although their masses will depend on $g_{YM}$.

The theory has 1/2-supersymmetric BPS solitons or 0-branes given
by
lifting self-dual Yang-Mills instantons in 4
Euclidean dimensions to soliton world-lines in 4+1 dimensions.
These
solitons have mass
$$M \propto {\vert n\vert \over g_{YM}^2}
\eqn\abc$$
where $n$ is the instanton number, so that these become light as
$g_{YM}$ becomes large.
A supersymmetric theory in 6 dimensions compactified on a circle
of
radius $R$
gives a tower of 1/2 supersymmetric multiplets of BPS
massive states with momentum $n/R$ in the circular dimension for
integers $n$, with masses
$$M \propto {\vert n\vert \over R}
\eqn\abc$$
 In [\roz], it was suggested that the light solitons in $D=5$
Yang-Mills
should be re-interpreted as Kaluza-Klein modes
 for a 6-dimensional theory on a circle of radius $R\propto
g_{YM}^2$.
 Theories in six dimensions with 16 supersymmetries can have
either
(1,1) supersymmetry or (2,0)
supersymmetry\foot{
$(p,q)$ supersymmetry in 6 dimensions has $p$ right-handed
symplectic Majorana-Weyl supercharges and $q$ left-handed ones},
but only (2,0) supersymmetry  allows $Spin(5)$ R-symmetry and
keeping
the same number of scalars, and this gives a unique candidate
for a strong coupling limit
that is consistent with the assumptions, which  is the
superconformal
(2,0)
theory. In the abelian case, this is the free field theory of the
(2,0) tensor multiplet in 6 dimensions. This multiplet consists of
 5 scalars $\phi^I$,
four 4-component symplectic Majorana-Weyl fermions,
and a 2-form gauge field
$B_{{MN}}$ satisfying a self-duality
condition $H=*H$ where $H$ is the field strength $H=dB$.
This theory has no dimensional coupling constants
(after rescaling the fields)
and is superconformal invariant
with superconformal group
$OSp^{*}(8/4)$. This has
  bosonic subgroup $SO^{*}(8)\times USp(4)=SO(6,2)\times Sp(2)$
which is the product of the conformal group and the R-symmetry
group, together with 32 fermionic generators.
Dimensional reduction on a circle of radius $R$
gives the $D=5$ abelian super-Yang-Mills theory with
action \ymact\ and coupling constant $g_{YM}^{2}=R$.

This leads to the conjecture that
the strong coupling limit of the
$D=5$ super-Yang-Mills theory
is an interacting 6-dimensional superconformal theory with (2,0)
supersymmetry
and $OSp^{*}(8/4)$ superconformal symmetry compactified on a
circle of
radius $R=g_{YM}^{2}$ [\roz-\note].
In the abelian case, it is easily verified that  dimensional
reduction of the (2,0) theory gives the $D=5$ theory, but
there are no local covariant interactions that can be written for
the tensor multiplet that  give the non-abelian $D=5$ interactions on
dimensional
reduction.
Nonetheless, there are arguments that such an interacting theory
in
$D=6$
should exist, based on the assumption that
M-theory is consistent.
For example, it arises as part of the world-volume theory
for a stack of M5-branes [\strom] and there is a decoupling limit
leaving precisely the (2,0) interacting superconformal
theory [\roz-\note].
The consistency of this limit and the existence of a non-trivial
quantum theory in six dimensions depends crucially on the
conformal
invariance of the (2,0) theory.
An important check on this interpretation is that
 when the original 4 spatial dimensions are compactified on $T^4$,
the theory has an $SL(5,\Z)$ invariance  [\roz,\brs], while for the
uncompactified
 theory in 4+1 dimensions there is evidence that the theory has 5+1
dimensional Lorentz
invariance  in the limit
 $g_{YM}\to \infty$.
The M5-brane realisation leads to a matrix theory construction of
the interacting theory in discrete light-cone gauge [\matten].

The interacting theory    then has some \lq M-interactions'
which
give rise to the 5-dimensional non-abelian interactions but do not
have a conventional field theory formulation in 6 dimensions.
It is not   clear whether the tensor multiplet provides the
correct
degrees of freedom for the interacting theory,
or whether something more exotic such as  string fields might be
needed [\note].
It was originally suggested that the interacting (2,0) theory might be a theory
of tensionless non-critical strings [\WitUSC],
but later it was proposed that it could be a  superconformal quantum field
theory
of a more conventional type (see e.g. [\seig]).

The scalars $\phi$ in the action \ymact\
have dimension one, but in 6-dimensional superconformal field
theory
the canonical dimension for scalars is two.
Then defining the dimension-2 scalars
$$\Phi^{I}={1\over g_{{YM}}^{2}}\phi^{I}
\eqn\abc$$
the 6-dimensional
action
$$
-\2 \int d^{6}x  (\pa \Phi)^{2}+\ldots
\eqn\abc$$
has no dimensional parameters but gives
the scalar part of \ymact\ on reducing on a circle  of
radius $R=g_{YM}^{2}$.
The limit leading to the (2,0) theory is
a strong coupling limit in which $g_{YM}^{2}\to \infty$ keeping
$\Phi^{I}={1\over g_{{YM}}^{2}}\phi^{I}$ fixed, so that the
scalars
$\phi$ are all scaled with $g_{YM}^2$. The (2,0) theory
has no dimensionful parameters and     scales
are set by the expectation values
of the scalars $\Phi$.

The
BPS states of the $D=5$ theory
include   the instantonic 0-branes, electrically charged 0-branes
and
magnetically charged
strings.
The electric  0-branes are  the W-boson multiplets with mass
$M\propto \phi
$ where $\phi$ is the magnitude of the
scalar expectation value, $\phi = |<\phi^I>|$,
and the BPS  strings have tension $\Phi= \phi/g_{YM}^2$.

The five-dimensional $N=2n$ superalgebra with automorphism group
$Sp(n)$ is
$$\eqalign{
\{Q_\alpha^a,Q_\beta^b\} &=\, \www^{ab}\big(
\Gamma^\mu C\big)_{\alpha\beta} P_\mu  +
\www^{ab} C _{\alpha\beta}K
\cr
& +
\big( \Gamma^\mu C\big)_{\alpha\beta} Z_\mu ^{ab} +
 C _{\alpha\beta}Z^{ab}  +\2 \big(
\Gamma^{\mu\nu }C\big)_{\alpha\beta}Z_{\mu\nu }^{ab}
\cr}
\eqn\fialpp$$
where $\mu,\nu=0,1,\dots , 4$ are spacetime indices, $\aa=1,\dots ,4$  are
spinor indices, $a=1,\dots ,N$ are
$Sp(n)$ indices,
$C _{\alpha\beta}$
is
the charge conjugation matrix and $\www^{ab}$ is
the symplectic invariant of
$Sp(n)$.
The supercharges $Q_\alpha^a$ are
symplectic Majorana spinors
satisfying
$$
(\bar Q)^\alpha_a= C^{\aa \bb} \www_{ab}
Q_\bb ^b
\eqn\real$$
For the $N=4$ super-Yang-Mills theory, the
5 central charges $Z^{ab}=-Z^{ba}$ with
$\www_{ab}Z^{ab}=0$
are proportional to the five electric charges
$q^{I}\propto \int tr (*F \phi^{I})$ and are carried by massive vector
multiplets.
There is no vector field coupling to the   singlet central charge
$K$,
which  is a  topological charge carried by
the instantonic 0-branes and is proportional to the instanton
number.
The spatial components
$Z_i^{ab}$  ($i,j=1,...,4$) of $Z_\mu ^{ab}$
are the  magnetic charges
  carried by the magnetically charged strings.
  The other charges on the right-hand-side of \fialpp\ are
  2,3 and 4-brane charges  [\GravDu]; for example, a vortex
solution
  in 2+1 dimensions   lifts  to a 2-brane in 4+1 dimensions with charge
$Z_{ij}$.

  There are two distinct  types of 1/2 supersymmetric BPS 0-brane
  representations [\falg]. The first has central charges  $K=0$ and  $Z^{ab}
  = i q\ss^3\otimes \ss^2$ for some charge $q$, breaking the R-symmetry from
$Spin(5)$
to $Spin(4)$,
     and is a massive vector multiplet with mass $M=|q|$.
  The second has $Z=0$ and $K\ne 0$ and
  is a massive self-dual tensor multiplet with mass $M=|K|$
[\falg]
  and preserves  the full $Spin(5)$ R-symmetry.
The
  W-bosons fit into massive vector multiplets
  and
  the instantonic 0-branes fit into massive self-dual tensor
multiplets.
  The limit in which the tower of
  instantonic 0-branes becomes massless
  is necessarily a (2,0) theory in 6 dimensions; this is because
the
  Kaluza-Klein modes for a (2,0) tensor theory fit into
  massive tensor multiplets while those for a (1,1) theory in 6
  dimensions would fit into massive vector multiplets.
 The central charge $K$ becomes the 6th component of momentum,
$P^M=(P^\mu,K)$,
while the central charges $Z^{ab}$
fit  together with the $D=5$ string charges   $Z_{\mm}^{ab}$ to
give
the $D=6$ string charges
$Z_{M}^{ab}=(Z_{\mu}^{ab}, Z^{ab})$.

The six-dimensional (2,0) theory
has
BPS strings coupling to the self-dual anti-symmetric tensor gauge
fields
of tension
$\Phi$ where $\Phi$ is the
the magnitude of the
  expectation value of $\Phi^{I}$.
  On dimensional reduction,
  the unwrapped strings
  give the magnetic strings in five dimensions
  of tension $ \Phi=\phi/g_{YM}^2$ while the wrapped strings
  give the W-bosons with
  mass $M\propto R\Phi=\phi$.
  Thus the BPS spectra match, and the strong coupling limit is the
  one in which $g_{YM}\to \infty$ while the tensions of the BPS strings are
kept fixed.

  At the origin of moduli space at which the scalar expectation
values
all vanish, all W-bosons become massless and the full gauge
symmetry
of the $D=5$ Yang-Mills theory is restored and, as $\phi \to 0$,
the
tensions of the
  magnetically charged strings approach zero.
  In the six dimensional theory, the tensions of the BPS strings
also approach zero as  $\Phi \to 0$.
The nature of the theory at such points is unclear.  For example,
one
guess might be that it could be   described by some kind of
string field theory, but it was argued in [\note] that such a
description
would  over-count the degrees of
freedom.

  These arguments should apply to 5-dimensional Yang-Mills
theories
 independently of how they are embedded in other theories.
 The (2,0) superconformal field theory is believed to be a
consistent
 quantum theory in its own right,
 and so can provide a complete formulation of the theory described
at
 weak coupling by an effective 5-dimensional Yang-Mills theory.
 Conversely, any theory whose effective description is given by
5-dimensional
 $N=4$ Yang-Mills theory plus other $N=4$ multiplets will, in the
 limit in which the Yang-Mills coupling becomes strong while the
 magnetic string tensions remain constant, be the six-dimensional
 (2,0) conformal field theory plus other supermultiplets.
 For example, the dynamics of a stack of $m$ parallel D4-brane is described by
 $D=5$, $N=4$ Yang-Mills theory with $SU(m)$ gauge symmetry plus a tower of
massive $D=5,N=4$
 multiplets, coupled to  a 10-dimensional bulk theory consisting
of supergravity
 plus an infinite number of massive multiplets.
 At strong coupling, the D4-branes become   M5-branes wrapped on a
 circle of radius $R=g_{YM}^{2}$
 whose  dynamics are described by the
 $D=6$ (2,0) conformal field theory theory,  plus a tower of massive
 $D=6$ (2,0)
 supermultiplets, coupled to  an 11-dimensional bulk theory whose
massless
 sector is 11-dimensional supergravity compactified on a circle of radius $R$.
 As another example, the heterotic string compactified
 on $T^{5}$
 is described by
 $D=5$, $N=4$ Yang-Mills theory plus $D=5$, $N=4$ supergravity
 plus a tower of massive $D=5,N=4$
 multiplets. These include massive string modes and Kaluza-Klein
modes.
 At strong coupling this becomes the
 six-dimensional (2,0) supersymmetric string theory
 obtained by compactifying type IIB string theory on $K3$
[\WitUSC], further
 compactified to five dimensions on a circle of radius
$R=g_{YM}^{2}$.
The 6-dimensional (2,0) theory at generic points in moduli space is (2,0)
supergravity coupled to (2,0) tensor multiplets, together with an infinite
number of massive (2,0) multiplets, with tensionless strings appearing at
special points in the moduli space [\WitUSC].

 In both cases, each massless vector multiplet in $D=5$ becomes a
 $D=6$ (2,0) tensor
 multiplet at strong coupling at generic points in moduli space, with
tensionless strings occurring at special points.
 It is natural to expect the same to happen for   $N=4$ vector
 multiplets in any $N\ge 4$ supersymmetric theory in five
dimensions,
 and in particular   that the   abelian $N=4$ vector multiplets occurring in
$D=5,N=8$
 supergravity theory   become six dimensional (2,0) tensor multiplets.

\chapter{$N=8$ Supergravity in Five
 Dimensions}

 Maximal supergravity theories have
 a local symmetry group $H$
 and a global symmetry
$G$ [\julia], broken to a discrete U-duality  subgroup
in the quantum theory [\HT].
The scalars take values in the coset space
$G/H$ and,  if the local $H$ symmetry is fixed by choosing the
  symmetric gauge, then a diagonal  R-symmetry subgroup  $H$ of
$G\times
H$ is manifest.
The  $N=8$ supergravity theory in five dimensions [\Cremmer]
has  local $H=Sp(4)=USp(8)$ symmetry and a global
$G=E_{6}$ symmetry.
In symmetric gauge, the $N=8$ supergravity multiplet
decomposes into the following representations of the
$Spin(3) \times Sp(4)$ little group:
$$
2^8=(5,1)+(3,27)+(1,42)+(4,8)+(2,48)
\eqn\abc$$
The   vector
fields transform as the {\bf 27} of $E_{6}$ while
the 42 scalars take values in
$E_{6}/Sp(4)$. The Einstein-frame action takes the form \act, with
$l$ the 5-dimensional Planck length. If the theory is obtained
from 11
dimensional supergravity compactified on a 6-torus of volume
$L^{6}$, then $l$
is given in terms of $L$ and the 11-dimensional Planck length
$l_{{11}}$ by
$l=l_{11}^{3}/L^{2}$.

The $N=8$ superalgebra has automorphism group
$Sp(4)$ and is given by \fialpp\ with $a,b=1,\ldots,8$. The
algebra with
scalar central charges only is
$$
\{Q_\alpha^a,Q_\beta^b\} =\, \www^{ab}\big(
 \Gamma^\mu C\big)_{\alpha\beta} P_\mu
 +
 C _{\alpha\beta}(
 Z^{ab}+\www^{ab} K)
\eqn\fialgo$$
 The 27 central
charges
$Z^{ab}$ satisfy $Z^{ab}=-Z^{ba}$,
$\www_{ab}Z^{ab}=0$,
and are the electric
charges for the 27 vector fields (dressed with scalars).
There are 28 central charges but only 27 vector fields, and
$K$ is not the conserved charge for any  gauge field.
However, on dimensional reduction from 5  to 4 dimensions, $K$
becomes
one of the 28 magnetic charges of $D=4,N=8$ supergravity, and
is the one coupling to   the gravi-photon field (i.e. the
electromagnetic field from the dimensional reduction of the
metric). U-duality requires that 1/2-supersymmetric states with $M=|K|$ occur
in the $D=4$ BPS spectrum, and   $K$ is quantized [\HT].
However, these $D=4$ states must arise from
1/2-supersymmetric states in $D=5$, carrying the central charge that gives the
Kaluza-Klein magnetic charge in four dimensions. They must then be
1/2-supersymmetric $D=5$ states
 with $M=|K|$, and   $K$ should be
quantized, so that
$K\propto n/l$   where $n$ is an  integer.

Thus BPS
states carrying the charge $K$ should be present in the $D=5$ spectrum.
A definition of the charge $K$
was given in [\GravDu], where it was shown that it indeed arises in the
superalgebra
\fialgo.
Examples of 1/2-supersymmetric supergravity solutions carrying the
 charge $K$ are metrics  of the form
 $N\times \R$ where $N$ is a self-dual gravitational  instanton [\GHaw] in
4
Euclidean dimensions and $\R$ is time, with all other fields
trivial [\GP,\GravDu].
If $N$ is the multi-Taub NUT
instanton, the solution has an interpretation as a multi
Kaluza-Klein
monopole space [\SGP] and $K$
is the Kaluza-Klein monopole charge.
If $N$ is an ALE space, the
solution can be interpreted as representing a set of 0-branes
or ALE-branes  [\GravDu].

The BPS representations in $D=5$  were classified in [\falg].
A non-zero value for $Z$ breaks the $E_6$ to a subgroup and
BPS states carrying the charge $Z$ will have masses dependent on
the
42 scalars, through standard BPS formulae, as in e.g. [\HTE].
However, if $Z$ is zero and $K$ is non-zero, the $E_6$ is unbroken
($K$ is an $E_6$ singlet) and BPS states with mass $M=\vert
K\vert$
  preserve half the supersymmetry and have a mass that is
independent of all the scalar fields (in Einstein frame).
Such states     have Einstein-frame mass
$$
M= {c\vert n \vert
\over l}
\eqn\mas$$
for some numerical constant $c$ and    integer  $n$.
  Comparing with the $D=4$ BPS spectrum shows that
states should occur for all values of the integer $n$.
For the gravitational instanton solutions,
  $n$ is a topological number (essentially the NUT charge or
instanton number) [\GravDu]. The mass in $D=5$ string frame is
$$
M_s= {c\vert n \vert
\over lg^{2/3}}
\eqn\abc$$
   with an unusual power of
the string coupling $g$.

The BPS spectrum is then expected to have  $M=|K|$ BPS states
with \mas\ for all $n$ which    become light as $l \to \infty$
and
the examples of [\roz,\Witten]   suggest  identifying them
with
Kaluza-Klein modes for a 6-dimensional theory compactified on a
circle of radius $R\propto l$.
Then the charge $K$ would become the 6th component of the
momentum
and these states becoming light would not break any of the
$E_6\times
Sp(4)$ duality symmetry.
Of course, this suggestion requires many checks.
However, if this limit does give a 6-dimensional theory,
it can be expected to give a theory with
32 supersymmetries and $E_6\times Sp(4)$   symmetry.
 These BPS states fit into massive $D=5$
multiplets containing
42 massive scalars and 27 massive self-dual 2-form fields
$B$, satisfying $dB=m*B$, in the {\bf 27} of $E_6$  [\falg] (see also section
8).
 This is precisely what is needed for the
Kaluza-Klein modes of a massless 6-dimensional theory with
27 self-dual 2-forms and 42 scalars.
Then if these instantonic 0-branes are to be interpreted as
Kaluza-Klein modes of a 6-dimensional theory with 32
supersymmetries, then the $D=6$
theory will have 27 self-dual 2-forms and 42 scalars, as demanded
in the introduction.

BPS states  in M-theory or string theory have associated supergravity solutions
which in some cases  are singular and in others are non-singular.
The 1/2-supersymmetric supergravity solutions
carrying the $K$-charge with $Z=0$ include the
multi-instanton solutions $N\times \R$.
The multi-instanton metric of [\GHaw]  has Dirac (or Misner) string
singularities, although the spin-connection is well-defined.  The space
  is asymptotic to
flat $\R^4$ (the curvature is finite and falls off asymptotically, even though
the metric is ill-defined).
A non-singular metric is obtained if  the charges of the sources are all equal,
and suitable
identifications are made to give an ALE space which is like
$\R^4/\Gamma$ asymptotically, for a discrete group $\Gamma$ (which is $\Z_n$
for the explicit solutions of [\GHaw]).
Without the discrete identifications, the solutions with various $K$ charges
are
asymptotic to $\R^4$ but with singular metrics.
However, as will be discussed in section 6,
these singularities can be avoided by using dual variables.
It will be assumed that the BPS states with $M=|K|$
can be viewed as 0-branes and   can be identified with Kaluza-Klein modes.

\chapter{Theories with 32 Supersymmetries in   Six Dimensions}

The considerations of previous sections have suggested looking for
a
conformal
theory in six dimensions with 32 supersymmetries, $E_{6}\times
Sp(4)$
symmetry,
27 self-dual 2-forms and  42 scalars.
In this section the various $D=6$ multiplets with
32 supersymmetries will be discussed.
In six dimensions, the $(p,q)$ superalgebra has
$p$ right-handed symplectic Majorana-Weyl supercharges and $q$
left-handed ones, with automorphism group
$Sp(p)\times Sp(q)$.
The massless representations
decompose into representations of the
little group $SU(2)\times SU(2)\times Sp(p)\times Sp(q)$,
and a   list of \lq physically acceptable'
representations (reducing to $D=4$ representations with spin not greater than
2) is given in
[\Strath].
For 32 supersymmetries, there are three possibilities,
(2,2),(3,1) or (4,0) supersymmetry (together
 of course with (1,3) and (0,4)).\foot{The possibility of (3,1) and (4,0)
multiplets in $D=6$ was
noted in [\ton].}
The (2,2) multiplet has the
$SU(2)\times SU(2)\times Sp(2)\times Sp(2)$ content
$$\eqalign{
&(3,3;1,1)+(1,3;5,1)+(3,1;1,5)+(1,1;5,5)+(2,2;4,4)
\cr &
+(2,3;4,1)+(3,2;1,4)+(2,1;4,5)+(1,2;5,4)
\cr}
\eqn\reptt$$
and is the maximal supergravity multiplet in 6 dimensions,
obtained by reducing
11-dimensional supergravity on a 5-torus.
The supergravity has duality symmetries $G=SO(5,5)$ and
$H=SO(5)\times SO(5)\sim  Sp(2)\times Sp(2)$, and the graviton
is in the (3,3;1,1) representation.

The (3,1) multiplet has the
$SU(2)\times SU(2)\times Sp(3)\times Sp(1)$ content [\Strath]
$$\eqalign{
&(4,2;1,1)+(3,1;6,2)+(1,1;14',2)+(2,2;14,1)
\cr &
+(3,2;6,1)+(2,1;14,2)+(1,2;14',1)
\cr}
\eqn\repto$$
The (3,1) multiplet
has 28 scalars in the $(14',2)$ representation
of $Sp(3)\times Sp(1)$,
and the unique coset structure that this is consistent with
is $${F_{4}
\over
Sp(3)\times Sp(1)}
\ek
suggesting that $G=F_{4}$ and $H=Sp(3)\times Sp(1)$ in this case;
it is   remarkable that the exceptional group $F_4$
plays a role here (in the non-compact form with maximal compact
subgroup
$Sp(3)\times Sp(1)$.) There are 14 vector fields and 12 self-dual 2-forms.

The (4,0) multiplet has the
$SU(2)\times SU(2)\times Sp(4)$ content [\Strath]
$$(5,1;1)+(3,1;27)+(1,1;42)+(4,1;8)+(2,1;48)
\eqn\repfo$$
The (4,0) multiplet is remarkably simple, and
has all the properties demanded in the introduction:
there are 27 self-dual 2-forms and 42 scalars,
it has an $Sp(4)$ R-symmetry and the spectrum is
  consistent with rigid
$G=E_{6}$  invariance (with the 2-forms transforming in the {\bf
27}).
Moreover,   the free theory based on this
multiplet
is a superconformally invariant theory, with conformal
supergroup $OSp^*(8/8)$. This has bosonic subgroup
$USp(8)\times SO^*(8)=Sp(4)\times SO(6,2)$ and 64 fermionic
generators.
Thus this multiplet satisfies {\it all} the conditions required
for a strong-coupling limit of the $D=5$ supergravity,
and a theory of this multiplet with M-interactions is the candidate dual
theory.

However,   both the (3,1) and (4,0) multiplets have no graviton
(which
would be in the
$(3,3)$ of $SU(2)\times SU(2)$)
but   instead have exotic tensor gauge fields.
The (4,0) multiplet has a gauge field whose physical degrees of
freedom are
in the $(5,1)$ representation  of $SU(2)\times SU(2)$ while the
(3,1) multiplet has a field in the $(4,2)$ representation.
These will be discussed in the next section, and the covariant gauge fields
with these degrees of freedom will be given.

\chapter{Exotic Tensor Gauge fields}

In discussing the representations under the $SU(2)\times
SU(2)$ little group for massless particles in 6 dimensions,
it is useful to introduce indices $A,B=1,2$ for the first $SU(2)$
and indices $A',B'=1,2$ for the second, together with $SO(4)$
vector
indices $i,j=1,...,4$, raised and lowered with the   metric
$\dd_{ij}$. The translation between the two types of index uses
the
sigma matrices
$\sigma _i^{AB'}$, and the indices $A,B'$ can be combined into a
4-component spinor index $\aa=(A,A')$.

The
 (4,0) supermultiplet has fields in the $SU(2)\times SU(2)$
  representations $(s,1)$ for $s=1,2,3,4,5$
 corresponding to rank $s-1$ totally symmetric tensors
 $\phi_{A_1...A_{s-1}}=\phi_{(A_1...A_{s-1})}$.
 The (1,1) is a scalar and the (2,1) is a chiral spinor.
 The (3,1) representation is a symmetric bi-spinor
$B_{AB}=B_{BA}$,
corresponding
 to a self-dual 2-form
$B_{ij}=-B_{ji}$,  $$B_{ij}= \2 \epsilon _{ijkl}B^{kl}
\eqn\abc$$
 while the (1,3) representation $B_{A'B'}$ is an anti-self-dual
2-form.

The (5,1) representation is a symmetric 4th rank $SU(2)$ tensor
$C_{ABCD}=C_{(ABCD)}$
which corresponds to a 4th rank tensor $C_{ij\, kl}$ of $SO(4)$
with the same algebraic properties as the Weyl tensor:
$$C_{ij\, kl}=-C_{ji\, kl}=-C_{ij\, lk}=C_{kl\, ij}; \qq C_{[ijk]l}= 0, \qq
C^k{}_{jkl}
=0
\eqn\weyl$$
In addition it is self-dual:
$$C_{ij\, mn}= \2 \epsilon _{ijkl}C^{kl}{}_{mn} = \2
C_{ij}{}^{mn}\epsilon_{klmn}
\eqn\abc$$
or $C=*C=C*$.

The (4,1) representation gives a spinor field
$\psi_{ABC}$
which translates to a spinor-valued 2-form
$\psi_{ijA}$ which is self-dual
$$\psi_{ijA}= \2 \epsilon _{ijkl}\psi^{kl}_{A}
\eqn\abc$$
This is
a spinor-valued self-dual   2-form field
$\psi _{ij }^\aa$ ($\psi=*\psi$) satisfying the chirality condition
$\ggg^0\psi_{ij} = \psi_{ij}$, or equivalently
$(\ggg ^j)^\bb{}_\aa\psi _{ij }^\aa=0$.

The (3,1) multiplet has a field
 $D_{ABCD'}=D_{(ABC)D'}$ in the (4,2) representation.
This corresponds to a 3rd rank tensor
$D_{ij\, k}$ satisfying
$$ D_{ij\, k}=-D_{ji\, k}, \qq D_{ij}{}^j=0, \qq
D_{[ij\, k]}=0
\eqn\deyl$$
In addition it satisfies the self-duality constraint
$$ D_{ij\, k}= \2 \epsilon _{ijlm}D^{lm}{}_k
\eqn\abc$$

It is straightforward to find the free covariant
 fields in six dimensions corresponding to these degrees of freedom.
The (3,1) representation arises from a 2-form gauge field $B_{MN}$
(where $M,N=0,1,...,5$) with the gauge invariance
$$ \dd B_{MN}= \pa _{[M} \ll _{N]}
\eqn\abc$$
and invariant field strength 3-form $H=dB$ satisfying the
self-duality constraint
$$
H_{MNP}= \6 \epsilon _{MNPQRS}H^{QRS}
\eqn\abc$$

The (5,1) representation
arises from a gauge field $C_{MN\, PQ}
$ with the algebraic properties of the Riemann tensor
$$ C_{MN\, PQ} =
-C_{NM\, PQ}
=
-C_{MN\, QP}
=
C_{PQ\, MN}, \qq C_{[MNP]Q}=0
\eqn\calg$$
and the gauge symmetry
$$ \dd C_{MN\, PQ} = \pa _{[M} \chi _{N]PQ} +\pa _{[P} \chi
_{Q]MN}-2
\pa _{[M} \chi _{NPQ]}
\eqn\delcis$$
with parameter $\chi _{MPQ}=-\chi _{MQP}$.
The invariant field strength is
$$G_{MNP\, QRS} ={1\over 36}(\pa_{M}\pa_{S}C_{NP\, RS}+\ldots)=
\pa _{[M} C_{NP]\,  [QR,S]}
\eqn\gis$$
so that
$$G_{MNP\, QRS} =G_{[MNP]\, [QRS]} =G_{QRS  \,   MNP}
\eqn\gisdfsds$$
and satisfies the self-duality constraint
$$G_{MNP\, QRS}
= \6 \epsilon _{MNPTUV}
G^{TUV}{}_{ QRS}
\eqn\gdu$$
or $G=*G=G*$.

The (4,2) representation
arises from a gauge field
$D_{MN\, P}$ satisfying
$$D_{MN\, P}=
D_{[MN]\, P},\qq
D_{[MN\, P]}=0
\eqn\dalg$$
with the gauge symmetry
$$ \dd
D_{MN\, P}
= \pa _{[M} \aa _{N]P} -\pa _{[M} \aa _{NP]}
\eqn\abc$$
with parameter $\aa _{NP} $.
The invariant field strength is
$$S_{MNP\, QR} =\pa _{[M} D_{NP]\,  [Q,R]}
\eqn\abc$$
and satisfies the self-duality constraint
$$S_{MNP\, QR}
= \6 \epsilon _{MNPTUV}
S^{TUV}{}_{ QR}
\eqn\sdu$$

Similarly, the covariant description of the spinor-valued 2-form
$\psi _{ij }^\aa$ in the
(4,1) representation
of $SU(2)\times SU(2)$ is a fermionic
 2-form
gauge field which is  also a Weyl spinor, satisfying
$\psi _{MN }^\aa=\psi _{[MN] }^\aa
$
where $\aa$ is now a 4-component $D=6$ Weyl spinor index.
The gauge symmetry is
$$\dd \psi _{MN } =
\pa_{[M}\ee_{N]}
\eqn\delyis$$
with parameter a spinor-vector $\ee_{N}^{\aa}$. The   field strength
$$
\chi_{MNP}^{\aa}=\pa_{[P} \psi _{MN] }^\aa
\eqn\abc$$
satisfies
the self-duality constraint
$$
\chi_{MNP}^{\aa}
= \6 \epsilon _{MNPTUV}
\chi^{TUV\aa}
\eqn\pdu$$
The field in the (4,0) multiplet is in the
(4,1;8) of $SU(2)\times SU(2)\times Sp(4)$
and   is a field
$\psi _{MN }^{\aa a}$  carrying  an
$Sp(4)$ index $a=1,\ldots,8$, and satisfying a
symplectic Majorana-Weyl reality condition.

The other representations of $SU(2)\times SU(2)$ occurring in the
$(p,q)$ supermultiplets have more conventional
covariant representations.
The (3,3) comes from a graviton $h_{{MN}}$, the (2,2) from a
vector
gauge field, the (3,2) from a chiral gravitino $\psi _{M}^{\aa}$,
the (2,1) from a chiral spinor and the (1,1) from a scalar.

\chapter{Electromagnetic Duality  }

In $D$ dimensions, an abelian 1-form gauge field $A_{\mm}$ can be
dualised
to an $n$-form $\ti A_{\mm_{1}\mm_{2}\ldots\mm_{n}}$
where
$$n=D-3
\eqn\abc$$
with field strength $\ti F=*F$ where $F=dA,\ti F=d\ti A$.
Electric charges
lead to  a well-defined potential
$A$ but $\ti A$ has Dirac string singularities.
Magnetically charged $n-1$
branes
give fields for which $\ti A$
is well-defined but $A$ has Dirac string
singularities.
   Magnetic  charges
   can be accommodated
   by allowing
   $A$ to be a connection
   for a
   non-trivial bundle, or by switching to a description in terms
of
   $\ti A$, or by allowing string
   singularities in $A$.
   The   generalisation of this duality     to non-abelian gauge theory is
problematic, and in particular
   one cannot write down covariant local non-abelian interactions
for
   $\ti A$ if $n>1$.

   This picture can be generalised to gravity; the results will be
   summarised here and developed in more detail elsewhere.
   Consider the free theory given by taking the $D$ dimensional
metric
   $$g_{\mm \nn}=\eta _{\mm \nn}
   +
   h_{\mm \nn}
   \eqn\abc$$
   and linearising Einstein's equations in the fluctuation
   $   h_{\mm \nn}
$ about the Minkowski metric $\eta _{\mm \nn}
   $. Similarly  to   the gauge theory case, in the free theory
   one can dualise one of the indices on  $   h_{\mm \nn}
$ to obtain a field
$$D_{\mm_{1}\mm_{2}\ldots\mm_{n}\, \nu}
=D_{[\mm_{1}\mm_{2}\ldots\mm_{n}]\, \nu}
\eqn\abc$$
or both indices to obtain a gauge field
$$
C_{\mm_{1}\mm_{2}\ldots\mm_{n}\,
\nu_{1}\nu_{2}\ldots\nu_{n} }
=
C_{[\mm_{1}\mm_{2}\ldots\mm_{n}]\,
[\nu_{1}\nu_{2}\ldots\nu_{n}] }
\eqn\abc$$
where again
$n=D-3$.
For example, the gauge field $C$ has
a field strength
$$G_{\mm_{1}\mm_{2}\ldots\mm_{n+1}\,
\nu_{1}\nu_{2}\ldots\nu_{n+1} }
\eqn\abc$$
defined by a formula similar to \gis,
related to the linearised Riemann curvature
$R_{\mm\nn\rr\ss}(h)$
by
$$G_{\mm_{1}\mm_{2}\ldots\mm_{n+1}\,
\nu_{1}\nu_{2}\ldots\nu_{n+1} }
=\4
\ee _{\mm_{1}\mm_{2}\ldots\mm_{n+1}\aa\bb}
\ee_
{\nu_{1}\nu_{2}\ldots\nu_{n+1} \gg \dd}R^{\aa\bb\gg\dd}
\eqn\abc$$
or $G=*R*$.
As in the gauge theory case, the
generalisation to the interacting theory is problematic, and one
cannot write covariant local gravitational interactions
for the dual fields $C$ or $D$.

These dualities are easily derived in physical gauge.
Introducing transverse coordinates
$i,j=1,...,D-2$, a vector field
has physical light-cone gauge degrees of freedom $A_i$ in
the vector  representation
of the little group $SO(D-2)$ and can be dualised
to an $n=D-3$ form
$$\ti A_{j_1...j_n}= \ee _{j_1...j_n i} A^i
\eqn\abc$$
which can then be identified with the physical degrees of freedom
of
an $n$-form gauge field $\ti A_{\mm_1...\mm_n}$, with gauge invariance
$\delta\tilde A=d \lambda$.
The physical degrees of freedom
are in equivalent representations of the little group and so $A$
and
$\ti A$ give
equivalent field theory representations of the  physical degrees
of
freedom.

A physical gauge graviton is a transverse traceless tensor $h_{ij}$
satisfying
$$
h_{ij}=h_{ji}, \qq h_i{}^i=0
\eqn\hid$$
One or both of the indices on $h_{ij}$ can then be replaced by
$n$ anti-symmetric indices to give a dual form.
Choosing $D=5, n=3$ for simplicity,
one can define
$$
D_{ij\, k} = \ee _{ijl} h^l{}_k
\eqn\abc$$
so that \hid\ implies the
conditions
$$ D_{ij\, k}=-D_{ji\, k}, \qq D_{ij}{}^j=0, \qq
D_{[ij\, k]}=0
\eqn\abc$$
or
$$C_{ij\, kl} =\ee_{ijm} \ee_{kln}h^{mn}
\eqn\abc$$
which then satisfies
$$C_{ijkl}=-C_{jikl}=-C_{ijlk}=C_{klij}; \qq C_{[ijk]l}= 0, \qq
C^k{}_{jkl}
=0
\eqn\abc$$
These are precisely the conditions \weyl,\deyl\ found previously
in another
context, and
covariant  gauge fields
$C_{\mm\nn\, \rr \ss}$ and $D_{\mm\nn \, \rr}$ which reduce to
$C_{ij\, kl} $ or $
D_{ij\, k} $ can be found as in the last section.
These then give three equivalent   field representations of the
same degrees of freedom, and the duality of gauge theory generalises to a {\it
triality} of
(linearised)
gravitational theories.

In gauge theory, electric charges couple naturally to
$A$ but give a dual potential $\tilde A$ with Dirac string singularities while
  magnetic sources couple to $\tilde A$
 but give an $A$ with string singularities.
This generalises to
  linearised gravity, in which there are three types of
fields, $h,C$ and $D$, and if a source     leads to  a non-singular
configuration for
one of the
three fields $h,C,D$, then
that configuration  can in general have string singularities when expressed in
terms
of one of the other two fields. It seems natural to suppose that sources for
all three    fields could
arise in the full theory.
This would mean that, when
formulated in terms of $h_{\mu\nu}$ alone, some string singularities  in
$h_{\mu\nu}$ should
be anticipated.

Consider in particular the BPS states carrying the charge $K$ that were
discussed in section 3.
Reducing to
  $D=4$ gives magnetic monopoles which have vector potentials $A$ with Dirac
string singularities and,
as $A$ appears explicitly  in  the Kaluza-Klein ansatz for the $D=5$ metric,
the $D=5$ metric will
also have string singularities.
In $D=4$, the string singularities can be avoided by using the dual vector
potential $\tilde A$,
and this appears
in the Kaluza-Klein ansatz for the $D=5$ dual field $D_{\mu\nu\, \rho}$, as
$D_{m 4\, 4} \sim \tilde A_m$ ($\mu,\nu=0,1,\dots , 4$ and
$m,n=0,1,2 ,3$). This suggests that in
 a dual formulation of gravity using $D_{\mu\nu\, \rho}$
instead of  $h_{\mu\nu}$, states carrying the charge $K$
could correspond to field configurations without Dirac string singularities.

 \chapter{
Dimensional Reduction }

The dimensional reduction of a 2-form gauge field
$B_{{MN}}$ from 6 to 5 dimensions gives
a vector field $A_{\mm}=B_{\mm 5}$
and a 2-form $b_{\mm\nn}$ (where $\mm,\nn=0,1,\ldots4$
and $M,N=0,1,\ldots 5$).
In 5 dimensions, a vector field is dual to a
2-form field $\ti A$, with
$d\ti A=*dA$.
If $B$ is self-dual in $D=6$, $dB=*dB$, then in 5 dimensions
$db=*dA$, so that $b=\ti A$ is the 2-form gauge field dual to the photon. Then
$A$ and $b$ are not independent and the theory can be written
in terms of $A$ alone.
Thus a self-dual
$B_{{MN}}$   reduces to a vector field $A$ in $D=5$. Moreover, the
self-duality condition implies that $A$ satisfies the source-free Maxwell
equations.

A 4th rank gauge field
$C_{MN\, PQ}$ with the algebraic properties \calg\
gives
on reduction
the fields
$$h_{\mm\nn}=C_{\mm 5\, \nn 5},
\qq
d_{\mm\nn\,\rr}=C_{\mm\nn\,\rr 5},\qq
c_{\mm\nn\,\rr\ss}
=C_{\mm\nn\,\rr\ss}
\eqn\abc$$
However, it was seen in the last section that in the linearised
$D=5$ theory a linearised graviton
$h_{\mm\nn}$ has dual
representations in terms of
fields $D_{\mm\nn\,\rr}$ or $C_{\mm\nn\,\rr\ss}
$.
The 4th gauge field
$C_{MN\, PQ}$
occurring in the $D=6$ (4,0) multiplet
satisfies the self-duality condition \gdu\
which implies that the fields
$
d_{\mm\nn\,\rr}$, $
c_{\mm\nn\,\rr\ss}$ from dimensional reduction should be
identified
with the duals $D_{\mm\nn\,\rr}$, $C_{\mm\nn\,\rr\ss}$
of the $D=5$ linearised graviton
$h_{\mm\nn}$, so that the reduced theory can be written in terms
of
$h_{\mm\nn}$
alone, and the reduction of the exotic $D=6$ tensor gauge field
$C_{MN\, PQ}$ gives
 a conventional
$D=5$ graviton field $h_{\mm\nn}$.
The relationship
$$ R_{\mu\nu\sigma\tau}= {1 \over
6}\epsilon_{\mu\nu\kappa\lambda\rho}S^{\kappa\lambda\rho}{}_{\sigma\tau}
\ek
between the linearised Riemann curvature $R$ and the field strength $S$ for $D$
and the identity
$$S_{[\mu\nu\rho\sigma]\tau}=0
\ek
together imply that $h_{\mu\nu}$ satisfies the linearised Einstein equation,
$R_{\mu\nu}=0$.

The reduction of a $D=6$ field
$D_{MN\, P}$
satisfying \dalg\
 gives
$$\eqalign{&
h_{\mm\nn}=D_{5(\mm \, \nn) },
\qq
d_{\mm\nn\,\rr}=D_{\mm\nn\,\rr }
\cr &
A_{\mm }=D_{\mm 5\, 5 },
\qq
b_{\mm\nn }=D_{\mm\nn\,5 }
}
\eqn\abc$$
and if it satisfies the self-duality constraint \sdu,
then the 2-form gauge field $b_{\mm\nn} $ is dual to the vector
field $A_{\mm}$ and
the field $
d_{\mm\nn\,\rr}$ is dual to $h_{\mm\nn}$, so that the reduction
of  $D_{MN\, P}$
gives
    a
  graviton plus a vector field in $D=5$.
 Thus whereas the reduction of a $D=6$ graviton $h_{{MN}}$ gives a
 graviton, a vector and a scalar, the reduction of
 $D_{MN\, P}$ gives   a
 graviton and a vector but no scalar, and the reduction of
 $C_{MN\, PQ}$ gives a graviton only.

Similarly,  the   reduction of the $D=6$ fermionic field
 $\psi _{MN }^{\aa a}$ satisfying the self-duality constraint
\pdu\
gives
 the $D=5$ gravitini
 $\psi _{\mm }^{\aa a}$ of  $D=5,N=8$ supergravity, satisfying the linearised
$D=5$ Rarita-Schwinger equation.

The reduction of the other fields
 in the (2,2),(3,1) and (4,0) $D=6$ multiplets
 is straightforward, and the dimensional reduction of
 all three $D=6$ multiplets give the same
 $D=5$ multiplet, the $N=8$ supergravity multiplet.
 Thus the dimensional reduction of the free $D=6$ field theories
of the (2,2),(3,1) and (4,0)   multiplets
 all give  the same theory, the linearised $D=5$, $N=8$
supergravity. This was of course to be expected; although there
are three multiplets with 32 supersymmetries in $D=6$, there is
only one in $D=5$ (with \lq spins' not greater than two).

If the interacting gauge or gravity theories can be expressed as interacting
theories of the dual variables $\tilde A$, $C$ or $D$ introduced in   the last
section, then  the interactions could not be of a standard type, but they could
be M-interactions.
Indeed, the reduction of the (2,0) theory to
5-dimensions is naturally expressed in terms of a 2-form potential $B$
(together with its dual vector potential) and the $D=6$ M-interactions  should
give an M-interacting theory of
a 2-form gauge field in $D=5$. However,  this  is dual to a theory of $A$ alone
with the standard Yang-Mills interactions.
In the same way, reducing the M-interacting (4,0) theory (if such a theory
exists)
would give an M-interacting  theory in $D=5$
in terms of gauge fields $C,D$ and $h$, which can be dualised to
a theory of $h$ alone with
the standard local gravitational interactions.
Thus  M-interactions should play a central role in the dual formulations of
gravity or gauge theory.

\chapter{  BPS States in Five Dimensions}

The full $N=2n$   superalgebra in 4+1 dimensions is
given by \fialpp\
 with
$a,b=1,\ldots,N$.
For $N=8$, the charges and the BPS states carrying them are as
follows [\GravDu].
The 27
 central charges $Z^{ab}$
are the electric charges for the
 27 abelian  vector fields and are carried by electrically charged
0-branes,
the
  27 $Z_i^{ab}$
  with spatial indices
$i,j=1,...,4$
  are the corresponding magnetic charges
  carried by magnetically charged strings,
  and
  the charge $K$  is
the gravitational  charge of [\GravDu],
carried by BPS instantonic
0-branes.
There are  also 2-branes
coupling to axionic scalars with charge
$Z^{ab}_{ij}$, domain wall 3-branes with charge $\hat
Z_{ijk}^{ab}=\ee_{ijkl}Z^{0l\, ab}$
and space-filling 4-branes with charge
$\hat Z_{ijkl}^{ab}=\ee_{ijkl}Z_0^{ab}$.

The massive particle representations with central charges
$Z^{ab},K$
were classified in [\falg].
The supercharges transform as a $(2,1;N) + (1,2;N)$ under the rest-frame
little
group
$SU(2)\times SU(2)\times Sp(n)$, giving
the \lq chiral' supercharges
$Q_{\pm}^{a}$ with $Q_{\pm}^{a}= \pm\Gamma^0Q_{\pm}^{a}$.
For BPS representations with mass saturating a bound in terms of
the central charges
$Z^{ab},K$
which are invariant under the action of $2r_{+}$ of the
2-component
supercharges
$Q_{+}$ and $2r_{-}$ of the 2-component supercharges
$Q_{-}$, so that the fraction of supersymmetry preserved is
$\nu = (r_{+}+r_{-})/2n$,
the states will   fit into a representation
of the supersymmetry algebra generated by
  the remaining $4p$ positive chirality
supersymmetries and $4q$ negative chirality ones, with
$p=n-r_{+}$,
$q=n-r_{-}$. This is the  $(p,q)$ massive $D=5$ supermultiplet of
[\falg],
and the dimensional reduction on a circle  of a massless
representation of  $(p,q)$
supersymmetry in 5+1 dimensions gives a Kaluza-Klein tower of
massive
multiplets in 4+1 dimensions, each of which is a $(p,q)$ massive
$D=5$ supermultiplet with the same $p$ and $q$.
The central charge $Z$ breaks  the $Sp(n)$ to a subgroup
and the representation can be decomposed into
representations of
$SU(2)\times SU(2)\times
Sp(p)\times Sp(q)$,
which is   also the little group for massless
representations
of the $(p,q) $ superalgebra in 5+1 dimensions.

Of particular interest are those BPS representations preserving
1/2
the supersymmetry.
For $N=4$, there are two distinct such massive representations,
the (2,0) (or
(0,2)) multiplet and the (1,1) multiplet.
The massive vector multiplet is a (1,1) multiplet while the
massive
tensor multiplet is a (2,0) multiplet.
 For $N=8$, there are
three massive representations preserving 1/2 the supersymmetry,
the (2,2),(3,1) and (4,0) representations, with
little group decompositions
given by \reptt,\repto\ and \repfo\ respectively [\falg].

Free field theory representations can
be
found from the massive Kaluza-Klein modes of the circle compactifications
of  the massless $(p,q)$
representations in
6 dimensions discussed in previous sections.
For example, the massive
(3,1) multiplet in 5 dimensions has a
(4,2;1,1) of
$SU(2)\times SU(2)\times
Sp(3)\times Sp(1)$,
which is carried by a
field $D_{\mm \nn\,\rr}=D_{[\mm \nn]\,\rr}$ with
$D_{[\mm \nn\,\rr]}=0$
satisfying the massive $D=5$ self-duality constraint
$$\pa _{[\ss} D_{\mm \nn]\,\rr}
=m\ee _{\ss \mm \nn \kk \ll }D^{\kk \ll}{}_{\rr}
\eqn\abc$$
with mass $m$.


BPS 0-brane states preserving 1/2 the supersymmetry in (2,2),(3,1)
and (4,0)
multiplets all occur in the $D=5$ supergravity theory.
Those in (2,2) multiplets all have
$K=0$ and $Z\ne 0$ and the supergravity solutions are electrically charged
black
holes, while those in (3,1) multiplets have both
$K$ and
$Z$ non-zero. Those in (4,0) multiplets have $Z=0$ and $K\ne 0$ and the supergravity
solutions
are    gravitational instantons lifted to 4+1 dimensions
(self-dual instantons are in (4,0) multiplets while anti-self-dual
ones are in (0,4) multiplets).

Compactification of a $(p,q)$ supersymmetric theory in 5+1
dimensions
on a circle of
radius $R$ gives a Kaluza-Klein tower of massive
$(p,q)$ $D=5$ supermultiplets of masses $n/R$ for all positive
integers
$n$.
Conversely,
a  limit of a five dimensional $N=8$ theory in which a
mass $M\to 0$ and a
tower
of massive $(p,q)$ multiplets (with $p+q=4$) of masses $nM$,
$n=1,2,3,\ldots$, becomes light could be a decompactification
limit in which
an extra dimension opens up to give a 6-dimensional theory with
$(p,q)$
supersymmetry.
In particular,
a 0-brane with mass $M=\vert K\vert\propto n/l_{}$ will fit into a
$D=5$ (4,0) multiplet. The (4,0) multiplet has a free-field
representation
with 42 massive scalars,
27 massive self-dual 2-form fields $B$ satisfying
$$dB=m*B\ek
 and a massive
field $C_{\mm\nn\, \rr \ss}$
satisfying
$$\pa _{[\tt}C_{\mm\nn]\, \rr \ss}=
m \ee _{\mm\nn\tt \aa \bb}C^{\aa\bb}{}_{
 \rr \ss}\ek
 If these 0-branes are to be interpreted as Kaluza-Klein modes for
a massless
6-dimensional theory compactified
on a circle of radius $R=l$, then that 6-dimensional theory
must be (4,0) supersymmetric and have 42 scalars, 27 self-dual
2-forms and a self-dual $C_{MN\, PQ}$ gauge field,
as before.

\chapter {BPS Branes in Six Dimensions}

The (4,0) theory
has 27 self-dual 2-forms and these couple to 27 self-dual BPS strings, or
more
precisely, to strings whose charges take values in a
27-dimensional
lattice.
In addition, there are 42 scalars and these can couple to BPS
3-branes in 6 dimensions.

The (4,0) superalgebra in six dimensions with central charges
is
$$\eqalign{
\{Q_\alpha^a,Q_\beta^b\} &=\, \www^{ab}\big( \Pi_+
\Gamma^MC\big)_{\alpha\beta} P_M +\big(\Pi_+
\Gamma^MC\big)_{\alpha\beta} Z_M^{ab}
 \cr
&
  + {1 \over  6}. \big(\Pi_+
\Gamma^{MNP}C\big)_{\alpha\beta}Z_{MNP}^{ab}
\cr}
\eqn\fialgss$$
where $\Pi_+$ is the chiral projector
$$
\Pi_{\pm}=\2 (1\pm \ggg ^{7})
\eqn\abc$$
The 27 1-form charges satisfy
$Z_M^{ab} =-Z_M^{ba} $, $Z_M^{ab} \www_{ab}=0$, while
$Z_{MNP}^{ab}=Z_{MNP}^{ba}$ and is a self-dual 3-form,
$$Z_{MNP}^{ab}=\6 \ee _{MNPQRS} Z^{ab\, QRS}
\eqn\zdu$$
The 27 charges $Z_i^{ab}$ with spatial indices $i,j=1,\ldots,5$
are the string
charges, and
$Z_{ijk}^{ab}$
are the 36  3-brane charges, while  $Z_0^{ab}$
are the charges for space-filling 5-branes. (Note that the
self-duality
condition \zdu\
implies that $Z_{{0ij}}$ are not independent charges.)

On dimensional reduction to 5 dimensions, the
charges decompose as
$$ Z_{M}^{ab}\to (Z_{\mu}^{ab},Z_{5}^{ab}=Z^{ab}),
\qq
P^{M}\to (P^{\mu},P^{5}=K),
\qq
Z_{MNP}^{ab} \to Z_{\mm\nn 5}^{ab}= Z_{\mm\nn}^{ab}
\eqn\abc$$
The extra component of momentum becomes the $K$-charge, so that
the 5-dimensional instantonic 0-brane is, from the $D=6$ viewpoint, a wave
 carrying momentum $P^{5}$ in the extra circular dimension.
 The electric 0-branes in $D=5$ arise from strings winding around
the
 6th dimension while the magnetic strings arise from unwrapped
 $D=6$ strings. The 2-branes and 3-branes in $D=5$ arise from $D=6$
 3-branes, while the space-filling  $D=5$ 4-branes come from wrapped $D=6$
  5-branes.
  As in the (2,0) case reviewed  in section 2, the strong coupling
  limit $l\to \infty$ should be one in which the tensions of the
magnetic
  strings are kept finite if the BPS spectrums are to match,   as
  will be discussed in the next section.

  \chapter {The Strong Coupling Limit of $D=5,N=8$ Supergravity}

 In the previous sections,   evidence has been accumulated to
 support the conjecture that $D=5,N=8$ supergravity (embedded in some
consistent theory)
with
5-dimensional
 Planck-length $l$ is a $D=6$ (4,0) superconformal field theory
 compactified on a circle of radius $R=l$, so that the strong
 coupling limit $l\to \infty$ is one in which an extra dimension
 decompactifies. From the five-dimensional perspective, an
infinite
 tower of BPS 0-branes in massive (4,0) multiplets carrying the
 central charge
 $K$ become massless in this limit, signalling the opening up of
an
 extra dimension.
 For M-theory compactified on $T^{6}$, this  gives
 a six-dimensional strong-coupling \lq phase'
 which is (4,0) supersymmetric and which contains the
  (4,0) superconformal field theory.

 For the free theories, it has been shown  that the
linearised
 $D=5$ supergravity with Planck length $l$ is given by   the dimensional
reduction of the
 (4,0) tensor theory, formulated as a  theory
 of free fields propagating in a fixed background space-time
 which is a product of 5-dimensional Minkowski space and a circle
of
 radius $R=l$. For the interacting $D=5$ supergravity, the
 conjecture requires that there be an interacting form of the
(4,0) tensor
 theory which is also superconformal and whose dimensional reduction gives the
full supergravity,
 although there are no local covariant interactions for the (4,0)
multiplet
 with this property.
 This requires that the
 6-dimensional theory have some magical form of M-interactions
 which give the non-polynomial supergravity interactions on
 reduction. It could be that these are some non-local or
non-covariant
 self-interactions of the (4,0) multiplet, or it could be that
other
 degrees of freedom might be needed;  one candidate might be
 some form of string field theory.
 However, if a strong coupling limit of the theory does exist that
 meets the requirements
 assumed here, then the limit must be  a (4,0) theory in
six
 dimensions, and this would predict the existence of
M-interactions
 arising from the strong coupling limit of the supergravity
 interactions. Although it has not been possible to prove the
 existence of such a limit, it is remarkable that there is such a
 simple candidate theory for the limit with so many properties in
 common with the (2,0) limit of the $D=5$ gauge theory.
 Conversely, if there is a 6-dimensional phase of M-theory which
has
 (4,0) supersymmetry, then its circle  reduction to $D=5$ must
give an $N=8$
 supersymmetric theory and the scenario described here should
apply.

 Linearised gravity has dual descriptions in terms of exotic
tensor
 fields propagating in a fixed background, and these play an
important
 role in the construction used here. The triality between the formulations of
gravity in terms of $h,C$ or $D$, together with
the
 duality of abelian gauge theories, could have an extension to
the
 interacting non-linear theories in which M-interactions play an
 important part. Such matters would need to be understood before
 questions of back-reaction and background independence can be
 addressed. However, it is intriguing that M-theory seems to be
 pointing   to a more general framework than Riemannian geometry,
and
 one which would be better suited to accommodate \lq magnetic
masses' in
 gravity.

Both the (2,0) and (4,0) theories are chiral and so are liable   to anomalies.
In both cases there are potential
gravitational and supersymmetry anomalies, so that the coupling to gravity or
supergravity is problematic. For the M5-brane
world-volume theory,  the coupling to gravity is such that  these anomalies are
cancelled against bulk effects through anomaly
inflow [\anom], while for the IIB theory compactified on $K3$, the anomalies
from the (2,0) tensor multiplets are cancelled by
those from the (2,0) supergravity multiplet.

There is   presumably some non-abelian M-symmetry in the M5-brane world-volume
theory that reduces to the non-abelian gauge
symmetry in five dimensions and in six dimensions reduces to the 2-form gauge
invariance $\delta B=d\lambda$ in the abelian case. A conventional gauge
symmetry in a chiral six-dimensional theory would typically be anomalous,
but in the case of the  M-symmetry of a theory with M-interactions,  the
supposed consistency of the theory suggests that such
problems are absent. For the (4,0) theory, the inability to couple to gravity
is not a problem:  it is already a theory of
gravity of sorts (at least it dimensionally reduces to supergravity) and so
there is no need to couple to another gravity; if
one did, there would be two gravitons on reduction to five dimensions. Indeed,
there is no  (4,0) supergravity to attempt  to
couple to. There are presumably   M-symmetries in the (4,0) theory giving rise
to the diffeomorphism symmetry and local
supersymmetry of the $D=5 $ supergravity
and which reduces to the symmetries \delcis,\delyis\ in the
  free (4,0) theory. It seems plausible that, as for the (2,0) theory, the M-symmetry
could be free from anomalies, and the
consistency of the limit considered here would seem to require  this.

The electrically charged  0-branes of $N=4$ gauge theory and $N=8$ supergravity
in $D=5$  are lifted to wrapped strings of the (2,0) and (4,0)
theories, respectively.  In both cases there   are points in the $D=5$  moduli
spaces or their boundaries at which 0-branes
become massless and as these points are approached, the tensions of certain
$D=6$ strings   tend to zero. The
(2,0) and (4,0) superconformal  theories are very similar and both might be
regarded as theories of tensionless strings, as in
[\WitUSC].   However, it is possible that both are in fact quantum field
theories, with the superconformal invariance playing
a crucial role in their consistency. That this should be the case for the (2,0)
theory has been suggested in e.g. [\seig]. The
possibility that the  (4,0) theory could exist as an interacting superconformal
field theory is remarkable, as this would be a
 quantum theory whose   compactification
  would give $N=8$ supergravity in 4 or 5 dimensions, together with massive
fields.

 In six-dimensional conformal field theory, the natural dimension
for
 scalars is 2, and so it is useful to define
 the 42 dimension-2 scalars
 $$ \Phi =\phi/l^{2}
 \eqn\abc$$
 in terms of the 42 dimensionless scalars $\phi$ of the $D=5$
 supergravity theory. Writing  the 6-dimensional theory without
 coupling constants requires using $\Phi$ instead of $\phi$,
 and, just as in the (2,0) theory, the appropriate limit in the linearised
supergravity theory is
 $l\to \infty$ while $\Phi $ is kept fixed.
 Thus the limit involves scaling all
 the scalars $\phi$ in the same way,
 and can be viewed as a conformal transformation on the $D=5$
moduli
 space, taking one towards the boundary of moduli space in every
 possible direction simultaneously.
 Such limits were not included in the analysis of [\Witten,\HStr].

 The dimension of $\Phi$ makes it difficult to write
 down covariant local interactions in six dimensions without any
 explicit dimensionful parameters but which reduce to the
interactions in
 5-dimensions that are non-polynomial in $\phi$.
 This has particular relevance
 for the tensions of branes in the theory.
 Choosing any particular scalar $\phi$ (which could be  a string
theory
 dilaton, or  could be given by the volume of the M-theory
6-torus in Planck units $e^{\phi}=L/l_{11}$, for
 example), then the tension of a given $p$-brane
 in the 5-dimensional theory will depend on the
 asymptotic value $\phi_0$ of
 $\phi$ in general and is
 typically
 given by a formula of the form (in Einstein frame and suppressing a constant
of proportionality and dependance on the other scalars)
 $$T_{p}=e^{\aa _p\phi_0}{1\over l^{p+1}}
 \eqn\tp$$
 for some   $\aa_p$, which will depend on the choice of
$\phi$
 and on the choice of   brane.  For example, the part of the $D=5$ supergravity
action involving
 $\phi$ and a particular vector field
 $A$ will be of the form
 $$S=\int d^{5}x \sqrt{-g}\left( {1\over l^{3}}R-
{1\over 4l}e^{2a\phi}F^{2}- {1\over 2 l^{3}}
(\pa \phi)^2+
\ldots \right)
\eqn\acta$$
for some $a$, and $A$ will  couple to an electric 0-brane solution
with
$\aa_0= a$ and a magnetic string with $\aa_1=-a$.
 The 2-form gauge field $\ti A$ dual to $A$ has field strength
 $H=e^{2a\phi}*F$, so that the $D=6$ 2-form gauge field
 $B$ which reduces to $A,\ti A$ must satisfy
a non-linear self-duality constraint
whose $\phi$ dependence is given by
$$ H=e^{2a\phi}*H
\eqn\hcon$$

 To lowest order in $\phi$, \tp\ can be written in terms of
 the asymptotic value $\Phi_0$ of $\Phi$ as
 $$T_{p}=\left( {1\over l^{2}} + \aa _p\Phi _0 \right)
 {1\over l^{p-1}}
 \eqn\tplin$$
 dropping non-linear terms.  For a magnetically charged $D=5$ string with
$p=1$, this has a
 well-defined limit $\aa_1 \Phi_0$ as $l\to \infty$,
 provided  $\Phi$ is kept fixed (i.e. $\phi ,l \to \infty$ keeping
 $\phi/l^{2}$ fixed).
 The $D=5$ string then lifts to a   BPS string in $D=6$ dimensions
with
 tension  $\aa_1 \Phi_0$.
 Compactifying on a circle of radius $R$, a
  wrapped string of tension
  $\aa _1\Phi+1/R^{2}$ gives a 0-brane of mass proportional to
   $-R\aa _1 \Phi+1/R $, which agrees with \tplin\ for $p=0$
   (taking into account
   the $\phi$ dependence of \hcon\ with $a=-\alpha_1$, so that $\aa_0=-\aa_1$).
      Thus to lowest order in $\alpha_p$ the 6-dimensional strings
give
   strings and 0-branes in $D=5$ with the appropriate tensions.

   It is interesting to ask how this could generalise to a
   non-linear theory with M-interactions.
   The tension of the BPS strings in six dimensions should depend on
   $\Phi$ but not on any dimensionful parameters.
One possibility is that
  the tension could be of the form
   $$T_{1}=  \aa \Phi_0 +\2 \aa^{2} (\Phi *\Phi)_0 + O(\aa^{3})
   \eqn\abc$$
   where $\Phi *\Phi$ is some \lq M-product', which
must be of dimension 2, even though the simple product $\Phi^{2}$
has
naive
dimension 4, and which tends asymptotically to $(\Phi *\Phi)_0$.
On dimensional reduction on a circle of radius $R$, this must give
the 5-dimensional string tension, which requires that
 $
 (\Phi *\Phi)_0$ reduces to
$\Phi_0^{2}R^{2}=\phi_0^{2}R^{2}l^{-4}$
in five dimensions (perhaps with small corrections).
 This would then  give  the correct result if $R=l$, and
furthermore wrapped strings then give the correct 0-brane mass.
This   suggests that the $D=5$ tension formula \tp\ should be modified  to an
expression involving a $*$ product, with the property that the standard
supergravity results are
recovered at low energies but the limit $l\to \infty$ can be taken in such a
way that both $\Phi$ and
the string tension remain finite.

The strings in 6-dimensions then have
tensions which depend on the scalars $\Phi$ in some
mysterious  way, but on compactifying on a circle of finite radius
$R=l$, they must give
strings and 0-branes whose tensions depend on $\phi$ in the
standard
way \tp\ at low energies. In particular, at points in the moduli space
of the  5-dimensional theory in which the masses of some
$0$-branes or tensions of  some
strings become zero, the tensions of corresponding  strings in
6-dimensions should   approach zero or infinity.
The various strong-coupling limits of the 5-dimensional string
theory
(obtained by compactifying M-theory on $T^{6}$) that were
analysed in [\Witten, \HStr] in which
0-branes become massless
all correspond to limits
of the (4,0) theory
in which some string tensions become infinite and others approach zero.

For example, consider the
limit in which the volume in Planck units of the $T^{6}$ becomes
large, $L/l_{11} \to \infty$.
In that limit the Kaluza-Klein modes, viewed as 0-branes with
charges in a
6-dimensional lattice, all become massless, signalling a
decompactification to 11 dimensions.
In addition, there are six 0-branes whose masses become infinite
(from wrapped M5-branes), six strings that become tensionless (partially
wrapped M2-branes)
and six strings whose tensions become infinite (from wrapped Kaluza-Klein
monopoles).
Despite the presence of strings whose tensions approach  zero,
this limit of the 5-dimensional theory is a decompactification limit to
M-theory in 11-dimensional Minkowski space [\HStr].
However, if one first takes the strong gravitational coupling
limit to obtain
the
6-dimensional (4,0) theory,  then
instead of 6 types of 0-branes becoming massless, there are 6
types of
strings (with charges in a 6-lattice) which are becoming
tensionless, and a further 6 types of string whose tensions are becoming
infinite. The interpretation is not clear here, and in particular whether it
makes sense to
think of this in terms of a further 6 dimensions.
This limit is closely related to certain  limits discussed  in [\HStr], where a
12-dimensional interpretation was  considered.

The maximal  supergravity theory in $D\le 11$ dimensions together with its BPS
states and their
excitations contain much of   non-perturbative string theory. For example,
one of the BPS strings in the $D=5, N=8$ theory is the fundamental
string
for type II string theory compactified on $T^{5}$, and  the  perturbative
string states arise from fluctuations of this string.
In [\HT], the possibility that the
full non-perturbative string theory might in some way be equivalent
to an effective supergravity theory plus its solitons was suggested.
An obvious drawback with this is that the supergravity theory does not appear
to give a well-defined quantum theory.
However, if the
(4,0) theory is a
consistent superconformal field theory, then this (with its BPS states) could
give a formulation
of M-theory, with
the   string modes and Kaluza-Klein modes all carried by   strings. One of the
BPS strings will give the perturbative type II string on compactification
to $D=5$, while others will give     Kaluza-Klein modes.
Compactifying to  $D=5$ and then going to various boundaries of the moduli
space
will give decompactifications to other vacua of M-theory, including
11-dimensional Minkowski space.

Alternatively, it could be that the (4,0) superconformal field theory is a
sector of
a larger $D=6$  (4,0) supersymmetric  theory arising from the strong coupling
limit of M-theory compactified on $T^6$.
If the (4,0) superconformal field theory is consistent in its own right, it
might then arise as a decoupling limit
of the full theory. However, many of the known degrees of freedom of M-theory
are already contained in the (4,0) superconformal field theory as BPS branes
and their excitations.
The issue is perhaps whether the BPS branes (and their excitations) can be
treated properly, possibly as   solitons of some sort,  in the
(4,0) superconformal field theory, or whether they should be regarded instead
as independent degrees of freedom  (e.g. some sort of string fields) that need
to be introduced in addition.
Similar issue arise in the (2,0) tensor theory and even in $N=4$
super-Yang-Mills theory in four dimensions, where an infinite number of BPS
states that are usually treated as solitons -- the magnetic monopoles and dyons
-- become massless at particular points in the moduli space [\HTE].

Limits similar to the one considered here in other dimensions
 and other theories will be considered in a separate publication.
 It seems that there could be many phases of M-theory
which do not have a conventional field theory description but
which involve M-interactions, and understanding these should be of
considerable importance in unravelling the secrets of M-theory.

\refout

\bye